\documentclass{aastex631}

\usepackage[separate-uncertainty=true]{siunitx}
\usepackage{amsmath}

\newcommand{\figurewidth}{0.7\textwidth}

\DeclareSIUnit\gauss{G}
\DeclareSIUnit\erg{erg}
\DeclareSIUnit\pc{pc}
\DeclareSIUnit\year{yrs}
\DeclareSIUnit\solarmass{M_{\odot}}

\newcounter{daggerfootnote}

\begin{document}


\title{On the Potential Galactic Origin of the Ultra-High-Energy Event KM3-230213A}

\correspondingauthor{M.~Breuhaus}
\email{km3net-pc@km3net.de;breuhaus@cppm.in2p3.fr}

\author{O.~Adriani}
\affiliation{INFN, Sezione di Firenze, via Sansone 1, Sesto Fiorentino, 50019 Italy}
\affiliation{Universit{\`a} di Firenze, Dipartimento di Fisica e Astronomia, via Sansone 1, Sesto Fiorentino, 50019 Italy}
\author{S.~Aiello}
\affiliation{INFN, Sezione di Catania, (INFN-CT) Via Santa Sofia 64, Catania, 95123 Italy}
\author{A.~Albert}
\affiliation{Universit{\'e}~de~Strasbourg,~CNRS,~IPHC~UMR~7178,~F-67000~Strasbourg,~France}
\affiliation{Universit{\'e} de Haute Alsace, rue des Fr{\`e}res Lumi{\`e}re, 68093 Mulhouse Cedex, France}
\author{A.\,R.~Alhebsi}
\affiliation{Khalifa University of Science and Technology, Department of Physics, PO Box 127788, Abu Dhabi,   United Arab Emirates}
\author{M.~Alshamsi}
\affiliation{Aix~Marseille~Univ,~CNRS/IN2P3,~CPPM,~Marseille,~France}
\author{S. Alves Garre}
\affiliation{IFIC - Instituto de F{\'\i}sica Corpuscular (CSIC - Universitat de Val{\`e}ncia), c/Catedr{\'a}tico Jos{\'e} Beltr{\'a}n, 2, 46980 Paterna, Valencia, Spain}
\author{A. Ambrosone}
\affiliation{Universit{\`a} di Napoli ``Federico II'', Dip. Scienze Fisiche ``E. Pancini'', Complesso Universitario di Monte S. Angelo, Via Cintia ed. G, Napoli, 80126 Italy}
\affiliation{INFN, Sezione di Napoli, Complesso Universitario di Monte S. Angelo, Via Cintia ed. G, Napoli, 80126 Italy}
\author{F.~Ameli}
\affiliation{INFN, Sezione di Roma, Piazzale Aldo Moro 2, Roma, 00185 Italy}
\author{M.~Andre}
\affiliation{Universitat Polit{\`e}cnica de Catalunya, Laboratori d'Aplicacions Bioac{\'u}stiques, Centre Tecnol{\`o}gic de Vilanova i la Geltr{\'u}, Avda. Rambla Exposici{\'o}, s/n, Vilanova i la Geltr{\'u}, 08800 Spain}
\author{L.~Aphecetche}
\affiliation{Subatech, IMT Atlantique, IN2P3-CNRS, Nantes Universit{\'e}, 4 rue Alfred Kastler - La Chantrerie, Nantes, BP 20722 44307 France}
% ...




\begin{abstract}

The KM3NeT observatory detected the most energetic neutrino candidate
ever observed, with an energy between \SI{72}{\peta\electronvolt} and \SI{2.6}{\exa\electronvolt}
at the \SI{90}{\percent} confidence level. The observed neutrino is likely of cosmic origin. In this article, it is investigated if
the neutrino could have been produced within the Milky Way. Considering the low fluxes of the Galactic
diffuse emission at these energies, the lack of a nearby potential Galactic particle accelerator in the direction of the event and
the difficulty to accelerate particles to such high energies in Galactic systems, we conclude that if the event is indeed cosmic, it
is most likely of extragalactic origin.

\end{abstract}






\section{Introduction}

Neutrinos are unique cosmic messengers, produced exclusively
through hadronic interactions. They allow us to probe regions of
the Universe that are otherwise inaccessible and to discern leptonic from
hadronic processes, which is usually very difficult to achieve unambiguously
with electromagnetic observations alone. These elusive particles have opened
new avenues in high-energy astrophysics research. Neutrino detectors such as AMANDA,
IceCube, ANTARES, and Baikal-GVD demonstrated the feasibility of neutrino telescopes and provided the first groundbreaking results in
the last decade, such as the measurements of the astrophysical neutrino flux
\citep{IceCube_2013_astrophysical_neutrinos, IceCube_2016_astrophysical_neutrinos,
IceCube_2024_allsky_diffuse, ANTARES_2018_cosmic_neutrinos, ANTARES_2024_cosmic_neutrinos, Baikal-GVD_2023_astrophysical_neutrinos}
or the detection of neutrinos from the Galactic Plane \citep{IceCube_2023_Galactic_diffuse}.
Also, the first likely associations of neutrinos with single sources,
such as the blazar TXS 0506+056 or the active galaxy NGC 1068, have been achieved
\citep{IceCube_2018_TXS0506+056, IceCube_2022_NGC_1068}.

The KM3NeT neutrino telescope
has detected the event KM3-230213A on the 13th of February 2023 \citep{KM3NeT_2025_VHE_event}.
The event has a median neutrino energy
of \SI{220}{\peta\electronvolt}, with 
a \SI{90}{\percent} confidence interval ranging from \SI{72}{\peta\electronvolt} to \SI{2.6}{\exa\electronvolt}.
This makes KM3-230213A the most energetic neutrino candidate event observed so far.
Furthermore, the expected flux of atmospheric neutrinos above \SI{100}{\peta\electronvolt} in KM3NeT is $1-\num{5e-5}$ events per year \citep{KM3NeT_2025_VHE_event},
indicating that this event was
quite unlikely to be produced by cosmic rays (CRs) interacting in the Earth's atmosphere. This detection marks a significant milestone in
neutrino astronomy.

The origin of the event is not clear. While it is with extremely high probability a cosmic event, many different
classes of objects are expected to produce neutrinos, such as active galactic
nuclei (AGN), supernova remnants (SNRs), star-forming regions, $\gamma$-ray
bursts and others. Also, large-scale emission from the Galactic Plane, from
cosmogenic neutrinos, or unresolved sources can contribute to the neutrino flux
measured on Earth. However, only a few types of sources are expected to be able to accelerate
particles to the required energies.

In this article, we explore the possibility that the event is of Galactic origin,
i.e. produced in the Milky Way, where potential production sites are astrophysical accelerators, gas targets and cosmic-ray collisions giving rise to a diffuse Galactic component. 
The position of KM3-230213A was determined to be
RA = \SI{94.3}{\degree} and Dec = \SI{-7.8}{\degree},
with containment radii of
$R(\SI{68}{\percent}) = \SI{1.5}{\degree}$, and
$R(\SI{99}{\percent}) = \SI{3.0}{\degree}$. The corresponding Galactic coordinates are
$l = \SI{216.1}{\degree}$ and $b = \SI{-11.1}{\degree}$.
Therefore, the event occurred far away from the Galactic Centre, and
\SI{\sim 11}{\degree} offset from the Galactic Plane.
We first search for potential gas targets
within the vicinity of the event in \autoref{sec:molecular_gas}, as a correlation with gas is expected in the hadronic collision neutrino production channel. 
We then discuss the expected neutrino
fluxes from the Galactic diffuse emission (\autoref{sec:galactic_diffuse})
and further search for possible nearby particle accelerators (\autoref{sec:single_sources}).
The investigation is followed by a search for potential $\gamma$-ray counterparts and
a derivation of upper limits on Galactic neutrino fluxes as a result of the non-detection of sources
by the HAWC observatory (\autoref{sec:gamma-ray_neutrino_limits}). 
We conclude in \autoref{sec:conclusion}, where we summarize and discuss our results.

\section{Molecular gas in the neutrino event vicinity}\label{sec:molecular_gas}
    In many environments in the Galaxy, the dominant hadronic production mechanism
    for neutrinos
    and $\gamma$-rays is the collision of CRs with other hadronic targets.
    To investigate whether there is an enhancement of molecular
    gas close to the KM3-230213A position, we used carbon monoxide (CO) emission maps from \cite{Dame_2001_CO_survey}.
    CO is a reliable tracer of molecular
    gas and is often used to estimate the total mass of molecular gas
    \citep[e.g.][]{Neininger_et_al_1998_CO}.

\begin{figure}
    \centering
    \includegraphics[width=\figurewidth]{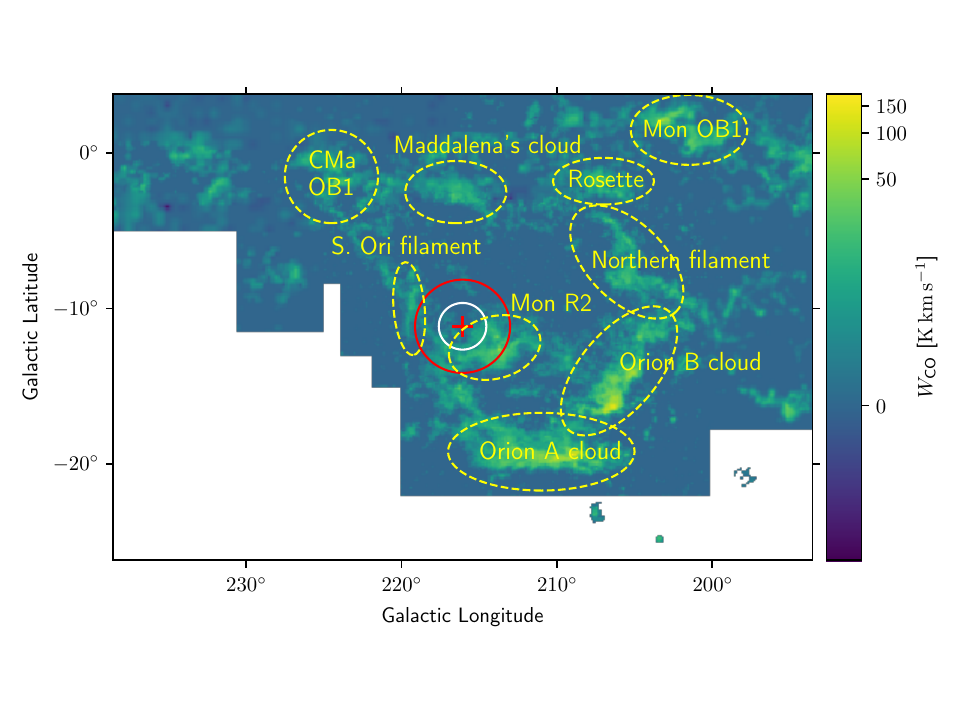}
    \caption{CO-map of the velocity integrated brightness temperature from \cite{Dame_2001_CO_survey} with the KM3-230213A event superimposed: The cross marks
    the event position, and the white and red circles are the \SI{68}{\percent}
    and \SI{99}{\percent} containment radii, respectively. The dashed ellipses mark the location of molecular
    clouds in the region.}
    \label{fig:CO_with_event}
\end{figure}

     The CO map of the velocity integrated brightness temperature together with the KM3-230213A
    position and the
    \SI{68}{\percent} and \SI{99}{\percent} containment radii is shown in \autoref{fig:CO_with_event}. The different
    molecular clouds in the region are also indicated. These are: CMa OB1, Maddalena's cloud, the
    Rosette molecular cloud, Monoceros OB1 (Mon OB1), the southern (S. Ori) and
    northern Orion filaments, Monoceros R2 (Mon R2), and the Orion A and Orion B
    clouds, with distances from the Earth between \SI{\sim 400}{\pc} and \SI{\sim 2.4}{\kilo\pc} \citep{Schlafly_2014_catalog_molecular_clouds}.
    As can be seen, no excess is observed directly at the
    KM3-230213A position.
    However, closeby there are the S. Ori
    filament, and the Mon R2 cloud, the latter being in overlap with the
    \SI{99}{\percent} containment radius of the neutrino event, thus it could be
    a potential target for CRs.
    Both the S. Ori filament and Mon R2 are relatively close to the
    Earth.
    
    The S. Ori filament is estimated to be at a distance of \SI{\sim 900}{\pc} and is potentially a 'bridge' between the Mon R2 and the CMa OB1 clouds
    \citep[see][and references therein]{Maddalena_et_al_1986}. Its mass estimated
    from CO observations is \SI{3e4}{\solarmass}, but estimates derived from
    the virial theorem or from the assumption of local thermodynamic equilibrium give
    larger values. It seems not to be a region of active star formation
    \citep{Maddalena_et_al_1986}.

    A commonly-adopted distance of Mon R2 from the Earth is \SI{830\pm 50}{\pc}
    \citep{Racine_1968, Herbst_Racine_1976}. It has a physical size of
    \SI{44}{\pc} $\times$ \SI{88}{\pc} \citep{Carpenter_2008_MonR2} and a total
    mass of \SI{\sim 9e4}{\solarmass} \citep{Maddalena_et_al_1986}. The molecular gas is not evenly distributed but exhibits a clumpy structure with a concentration towards the centre \citep[e.g.][]{Maddalena_et_al_1986}. Assuming a uniform distribution over a region of \SI{44}{\pc} $\times$ \SI{88}{\pc} $\times$ \SI{44}{\pc}, one obtains an average number density of \SI{\sim 20}{\per\cubic\centi\meter}. However, this estimate disregards the spatial shape of the cloud and from the CO maps one can infer that some fraction of the \SI{44}{\pc} $\times$ \SI{88}{\pc} diameter does not contain mass. In the densest regions of the cloud, number densities are likely to reach several hundreds of particles per \si{\cubic\centi\meter}. This is much larger than the average Galactic interstellar medium density of \SI{\sim 0.2}{\per\cubic\centi\meter} from the density model of \cite{Ferriere_1998_density_model, Ferriere_et_al_2007_density_model} at the same location.
    In contrast to the S. Ori filament, Mon R2 is a site of ongoing star formation
    \citep{Jiang_et_al_2024_MonR2_star_formation}. Since the Mon R2 cloud
    overlaps with the \SI{99}{\percent} containment radius of KM3-230213A,
    in principle, it could serve as a target for neutrino production.

\section{Potential cosmic ray sources}\label{sec:CR_sources}
    A big challenge for a potential Galactic origin of KM3-230213A is its extreme
    energy. In the case of hadronic collisions,
    the interacting CR needs to have at least an energy of several hundreds of
    \si{\peta\electronvolt}, assuming the lower limit for the neutrino energy of \SI{72}{\peta\electronvolt}. Thus it must be originating from sources contributing to the CR spectrum in the transition region between the ``knee" and the ``ankle" features. Many Galactic sources will fail to accelerate particles even to
    \si{\peta\electronvolt} energies. This is also reflected by the interpretation of the knee in the CR spectrum being caused by a rigidity-dependent
    cutoff of a Galactic source population \citep{Hoerendal_2003}.
    The probability for this event being of Galactic origin is reduced even more taking into account the distance from the Galactic Centre
    and the offset of \SI{11}{\degree} from the Galactic Plane. In this section, we will explore potential CR sources, first for the
    case of the diffuse emission, and later for other known Galactic
    $\gamma$-ray sources.

\subsection{Diffuse Galactic emission}\label{sec:galactic_diffuse}
    The diffuse Galactic emission is produced by CRs propagating in
    the Milky Way and interacting with other nuclei in the Galaxy. The CR spectrum
    measured at Earth extends up to energies of several \SI{e20}{\electronvolt},
    although the most energetic CRs are expected to have extragalactic origin \citep{Pierre_Auger_2024_CR_dipole}. Many different models for the Galactic diffuse emission were developed. These models differ, for example, in the considered CR composition, the diffusion properties of CRs in the Galaxy, and the distribution and injection of CRs from the sources. In particular, this modifies the subsequent CR distribution within the Galaxy in terms of total CR flux and potential spectral changes. Finally, these models use different $\gamma$-ray and CR datasets to obtain their estimates. The resulting $\gamma$-ray and neutrino emission is non-isotropic and follows the gas distribution of the Milky Way. We are specifically interested in the predicted emission at the location of KM3-230213A. To investigate how this localized flux compares to the overall emission, we use the model in \cite{Breuhaus_2022_hadronic_interactions} as an example case, as it was also focused on reproducing the $\gamma$-ray fluxes at hundreds of \si{\tera\electronvolt}, and compare the results later with predictions from the KRA$_{\gamma}^5$ model from \cite{Gaggero_et_al_2015_CR_model} and the newer KRA$_{\gamma}$ min. and max. models from \cite{Luque_et_al_2022_CR_models}.
    
    The model from \cite{Breuhaus_2022_hadronic_interactions} is tuned on $\gamma$-ray
    data from Tibet-AS$\gamma$ \citep{Tibet_2021_Galactic_diffuse} and ARGO-YBJ
    \citep{ARGO_YBJ_2015_Galactic_diffuse} for  Galactic longitudes $\SI{25}{\degree} \leq l \leq \SI{100}{\degree}$ and Galactic latitudes with $|b| \leq \SI{5}{\degree}$. Recently, the Galactic diffuse $\gamma$-ray emission was also observed by LHAASO \citep{LHAASO_2023_Galactic_diffuse}, hence we additionally tuned the model to these data only.
    The models are compared to the all-sky diffuse flux of KM3-230213A.

\begin{figure}
     \centering
     \includegraphics[width=\figurewidth]{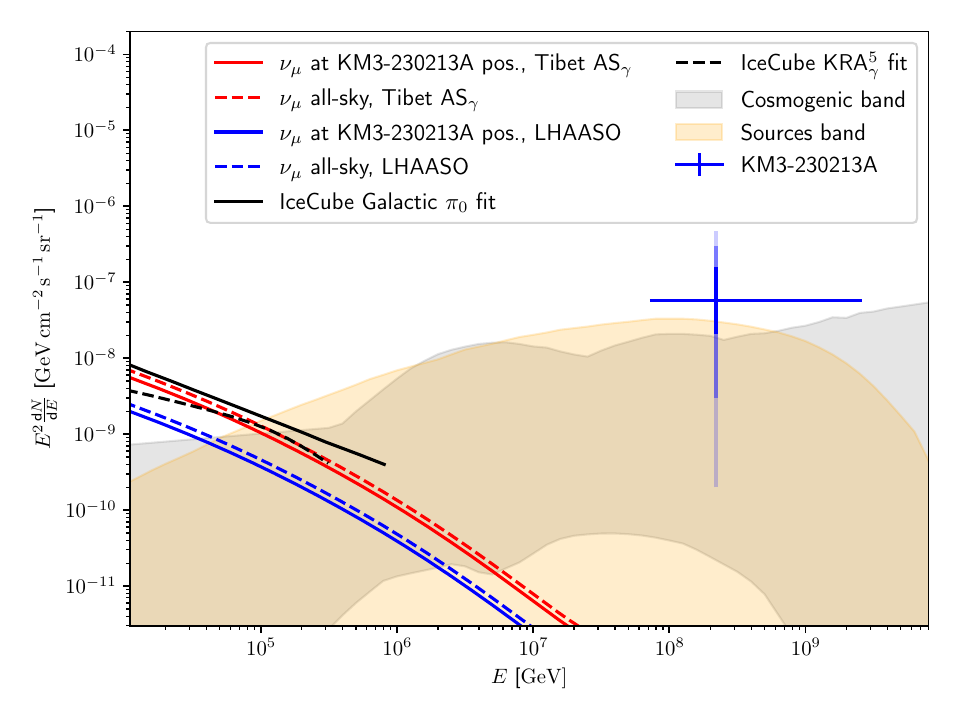}
     \caption{Expected neutrino fluxes from Model A of \cite{Breuhaus_2022_hadronic_interactions} compared to the flux derived from KM3-230213A (blue data point, the different blue shades represent the 1, 2 and \SI{3}{\sigma} Feldman-Cousins confidence intervals, respectively, \cite{KM3NeT_2025_VHE_event}). Dashed colored lines represent the total angle-averaged Galactic neutrino fluxes
        and solid lines the flux at the position of KM3-230213A. Models
        tuned on Tibet-AS$\gamma$ data are shown in red and the ones tuned on LHAASO data in blue.
     We also show measurements from IceCube of the Galactic diffuse neutrino flux for different models \citep[black lines][]{IceCube_2023_Galactic_diffuse} and the 
         bands of the cosmogenic neutrino flux and individual extragalactic sources shown in \cite{KM3NeT_2025_VHE_event}.}
    \label{fig:galactic_diffuse} 
    \end{figure}

\begin{figure}
    \centering
    \includegraphics[width=\figurewidth]{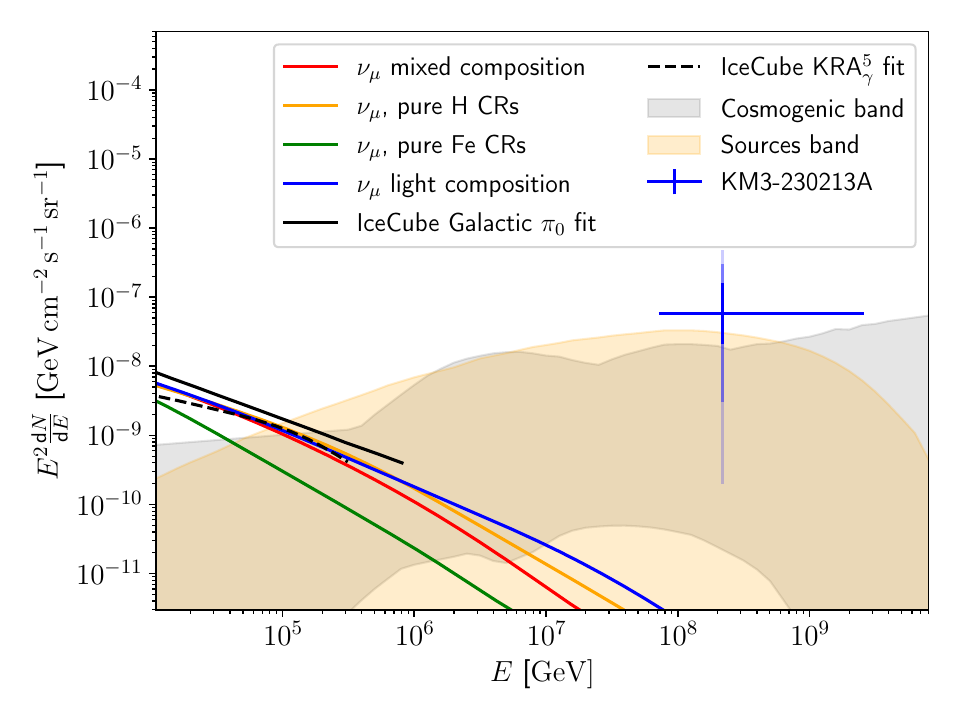}
    \caption{Expected neutrino fluxes from Models of \cite{Breuhaus_2022_hadronic_interactions} for different CR compositions compared to the flux derived from KM3-230213A (blue data point, the different blue shades represent the 1, 2 and \SI{3}{\sigma} Feldman-Cousins confidence intervals, respectively, \cite{KM3NeT_2025_VHE_event}). Model A is shown in red, the pure hydrogen and pure iron cases are shown in orange and green, and the blue line is the model with mixed composition below \si{\peta\electronvolt} energies changing to a pure hydrogen spectrum above for maximum neutrino emission.
     IceCube measurements of the Galactic diffuse neutrino flux for different models \citep[black lines][]{IceCube_2023_Galactic_diffuse} and the 
         bands of the cosmogenic neutrino flux and individual extragalactic sources from \cite{KM3NeT_2025_VHE_event} are also shown.}
    \label{fig:galactic_diffuse2}
    \end{figure}

    The results are shown in \autoref{fig:galactic_diffuse} and \autoref{fig:galactic_diffuse2}, where a comparison of the following neutrino fluxes is presented: the all-sky diffuse KM3-230213A flux \citep{KM3NeT_2025_VHE_event},
    measurements of neutrinos from the Galactic disk by IceCube
    for the $\pi^0$ and
    the KRA$_{\gamma}^5$ templates \citep{IceCube_2023_Galactic_diffuse}, and theoretical predictions for the cosmogenic neutrino flux and extragalactic sources taken from \cite{KM3NeT_2025_VHE_event}.
    The neutrino fluxes at the position of KM3-230213A and the
    all-sky diffuse Galactic neutrino flux from Model A  of \cite{Breuhaus_2022_hadronic_interactions} are shown in \autoref{fig:galactic_diffuse},
    tuned to data from Tibet-AS$_{\gamma}$ and to measurements from LHAASO
    of the inner Galaxy region ($|b| < \SI{5}{\degree}$, $\SI{15}{\degree} < l < \SI{125}{\degree}$). We note that although the LHAASO region of the inner Galactic Plane overlaps with the one from Tibet AS$_{\gamma}$ and ARGO-YBJ, it is not covering the same fraction of the sky. The model tuned on LHAASO data leads to lower predicted neutrino emission. This might be caused by the different source masking by Tibet AS$_{\gamma}$ and LHAASO and it reflects uncertainties in the measurements of Galactic diffuse emission by current $\gamma$-ray observatories. As can be seen, at energies above several tens
    of \si{\peta\electronvolt}, the Galactic diffuse fluxes are well below the KM3-230213A flux as well as the bands of the 
    cosmogenic neutrino flux and other extragalactic sources, making the association of KM3-230213A with a Galactic diffuse origin unlikely. Although the KM3-230213A flux is computed with the assumption of the emission being isotropic, which is not the case for the diffuse Galactic emission, using a non-isotropic template would not change the result significantly. 

    The exact composition of the CRs detected on Earth above \si{\peta\electronvolt}
    energies is not known and it can impact the resulting neutrino emission.
    To compare different emission scenarios, different compositional models from \cite{Breuhaus_2022_hadronic_interactions} are shown in
    \autoref{fig:galactic_diffuse2}.
    All of them are matched with the Tibet-AS$_{\gamma}$ and ARGO-YBJ data, resulting in higher
    fluxes than when matching to data from LHAASO.
    To explore the maximum possible neutrino flux at higher energies, we also show
    a case which was previously not in \cite{Breuhaus_2022_hadronic_interactions}:
    the composition of the first Galactic component from Model A in \cite{Breuhaus_2022_hadronic_interactions}
    dominating below \si{\peta\electronvolt} energies is kept as in the original model. All other CR contributions are assumed to be hydrogen above \si{\peta\electronvolt} energies, matching the all-particle CR flux. Therefore, above the energies where we can not measure the individual CR composition directly with satellite observations, the CRs are assumed to have the lightest possible composition. This is, as in the
    pure hydrogen and the pure iron case, not expected to be a realistic situation but it produces the
    maximum possible neutrino fluxes above \si{\peta\electronvolt} energies. The lower energies, which are
    matched to the $\gamma$-ray data, still reproduce the direct CR measurements
    on Earth. Even in this case, the diffuse Galactic emission
    is not expected to dominate over the cosmogenic fluxes.

\begin{figure}
        \centering
        \includegraphics[width=\figurewidth]{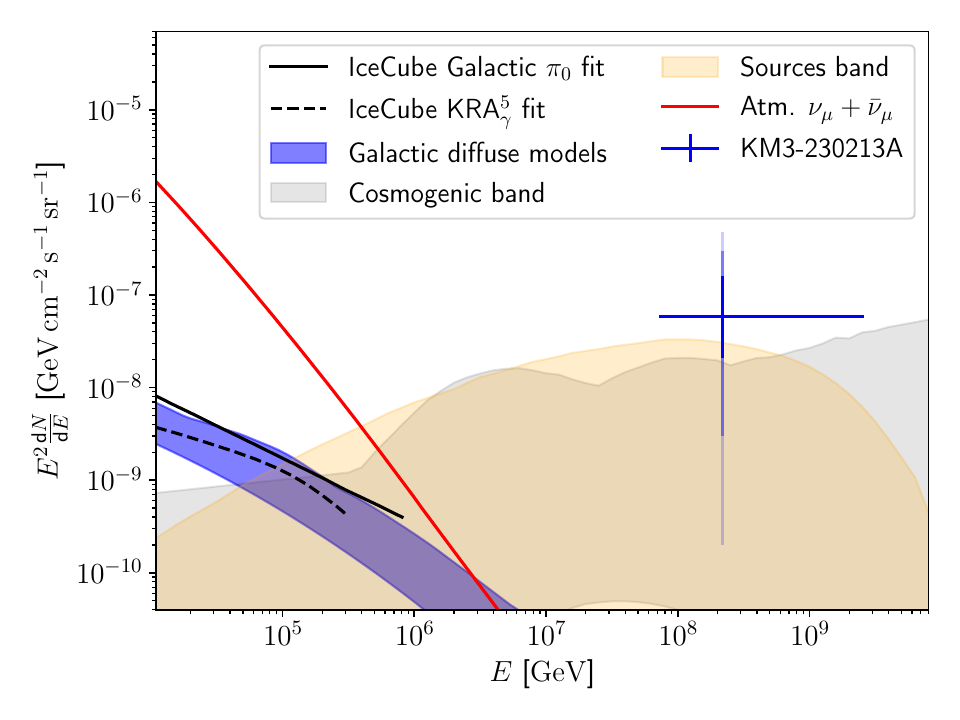}
        \caption{Expected fluxes from different models of the diffuse Galactic emission (blue band, see text for details) compared to measurements of the KM3-230213A all-sky diffuse flux from \cite{KM3NeT_2025_VHE_event} (blue data point, the different blue shades represent the 1, 2 and \SI{3}{\sigma} Feldman-Cousins confidence intervals, respectively.)
        IceCube measurements of the Galactic diffuse neutrino flux for different models are also shown \citep[black lines][]{IceCube_2023_Galactic_diffuse}, as well as bands of the cosmogenic neutrino flux and individual extragalactic sources as in \cite{KM3NeT_2025_VHE_event}.
 The zenith-angle averaged expected flux of atmospheric muon- and anti-muon neutrinos from \cite{Honda_2007_atm_neutrinos, Enberg_2008_prompt_atm_neutrinos} is shown in red.}
        \label{fig:diffuse}
    \end{figure}

Other models of the diffuse Galactic emission as in \cite{Breuhaus_2022_hadronic_interactions} give similar results. To illustrate this,
\autoref{fig:diffuse} shows the range of all-sky neutrino fluxes from the models of \cite{Breuhaus_2022_hadronic_interactions}, as well as the KRA$_{\gamma}^5$ model from \cite{Gaggero_et_al_2015_CR_model} and the newer KRA$_{\gamma}$ min. and max. models from \cite{Luque_et_al_2022_CR_models}. Due to the steep spectral index of the diffuse CRs, the resulting fluxes above tens of \si{\peta\electronvolt} are negligible compared to other sources. Below \si{\peta\electronvolt} energies the atmospheric muon- and antimuon neutrino flux dominates over the average diffuse Galactic neutrino flux. Above $\sim\si{\peta\electronvolt}$ energies, the diffuse Galactic emission is negligible compared to other expected neutrino fluxes, but it is still dominating over the atmospheric neutrinos \citep{KM3NeT_2025_VHE_event}.

By adopting a simple model for the Galactic density distribution at the location of KM3-230213A, such as the one from \cite{Ferriere_1998_density_model, Ferriere_et_al_2007_density_model}, the flux per steradian is reduced by \SI{\sim 20}{\percent}. This difference is smaller than changes caused by uncertainties in the CR models. The neutrino emission arising from dense target clouds, as Mon R2, is however expected to be enhanced. Estimations of the diffuse flux from Mon R2 were for example computed in \cite{HAWC_2021_MonR2}. Assuming that the emission is evenly distributed over an area of \SI{14}{\square\degree} \citep{Maddalena_et_al_1986}, enhancements of around one order of magnitude are possible compared to Model A from \cite{Breuhaus_2022_hadronic_interactions}. The material within Mon R2 is in fact not evenly spread out but rather concentrated in different clumpy regions \citep{Carpenter_2008_MonR2}. This will produce regions with both lower and higher $\gamma$-ray fluxes compared to computations assuming a homogeneous gas distribution. However, enhancements of many orders of magnitude, which would be needed for the diffuse Galactic neutrino flux to dominate over the cosmogenic neutrino flux or other extragalactic sources, are not possible. The diffuse Galactic neutrino flux is therefore negligible, both if it is produced in the diffuse interstellar medium and in the clouds investigated here.

\subsection{Single sources}\label{sec:single_sources}
    While the expected neutrino emission from the cloud Mon R2 caused by interactions of the diffuse Galactic CRs would not reproduce the flux inferred from the KM3-230213A measurement, the case in which a beam of accelerated particles from a nearby accelerator produces a larger emissivity can not be excluded.
    The challenge to explain the KM3-230213A neutrino with a single Galactic source comes from i) the presence of a possible accelerator, that can
    operate until the required energies, and ii) the gas target.
    A possible target for the neutrino event is the Mon R2 cloud (see \autoref{sec:molecular_gas}). An accelerator could also be located away from the
    emission site with its accelerated particles diffusing into Mon R2. However, if it is located
    too far away, the density of accelerated particles will be low. Additionally, if located at a similar
    distance to another cloud such as Orion A, the collisions with particles
    in this other cloud will also produce $\gamma$-rays and neutrinos. Depending on the
    gas densities, the emission from the second cloud can be comparable to or even higher
    than the emission from Mon R2. A source outside the Mon R2 cloud
    would produce $\gamma$-ray emission within the whole cloud and therefore Mon R2
    should appear bright in $\gamma$-rays. Such a scenario is
    inconsistent with the non-detection of Mon R2 by HAWC
    \citep[][see \autoref{sec:gamma-ray_counterparts}]{HAWC_2021_MonR2}. An optically thick source in $\gamma$-rays which is
    bright in neutrino emission can therefore be only located inside Mon R2, with
    the neutrinos produced in a rather small region with strong radiation fields.
    In the following, different potential source types are investigated.

\textbf{Supernova remnants (SNRs):}
Although SNRs might only be able to accelerate particles until \si{\peta\electronvolt}
energies in their earliest phases shortly after the supernova explosions
\citep{Lagage_Cesarsky_1983_SNR_CR_acceleration, Bell_2013_SNR_CR_acceleration, Marcowith_2018_SNR_CR_acceleration},
it was argued recently that SNRs in massive star clusters might be able to
accelerate CRs up to energies of hundreds of \si{\peta\electronvolt}
\citep{Vieu_2023_SNRs_in_SFR}. Even such an extreme scenario would only be able to produce neutrinos with energies of several tens of \si{\peta\electronvolt}, which is not sufficient to explain a neutrino with an energy much larger than the low end of the \SI{90}{\percent} confidence interval of KM3-230213A of \SI{72}{\peta\electronvolt}.

\begin{figure}
 \centering
 \includegraphics[width=\figurewidth]{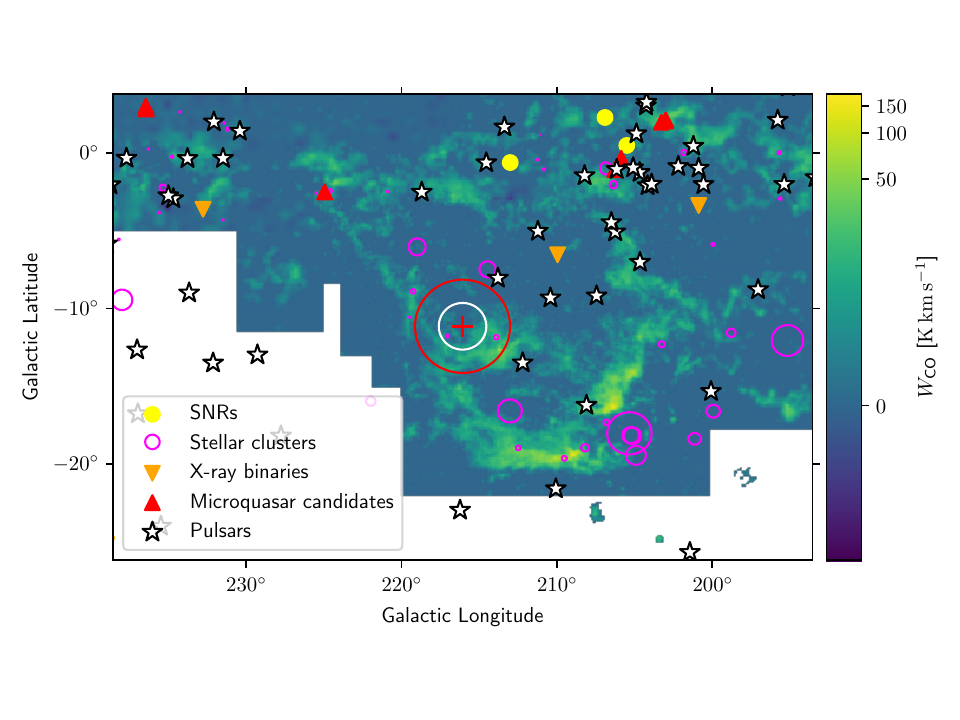}
 \caption{Different known potential CR accelerators in the region around KM3-230213A:
 The yellow dots represent SNRs from Green's SNR catalogue \citep{Green_2024_SNR_catalog}, and the magenta circles show the young stellar clusters from \cite{Celli_2024_young_stellar_clusters}, the sizes of the circles correspond to the cluster sizes. X-ray binaries and microquasars are displayed with the orange and red triangles \citep{Neumann_2023_Erosita_X-ray-binaries, Avakyan_2023_Erosita_X-ray-binaries}, and pulsars from the ATNF pulsar
 catalogue \citep{Manchester_2005_ATNF_pulsar_catalog} as white stars with black edges. The background is the velocity-integrated CO brightness temperature from \cite{Dame_2001_CO_survey} which traces the molecular gas.}
 \label{fig:sources}
\end{figure}

We compared the positions of known SNRs from the latest version of Green's SNR
catalogue \citep{Green_2024_SNR_catalog} with KM3-230213A (see \autoref{fig:sources}). No SNR close to KM3-230213A is known. Given the difficulties of accelerating CRs to the required
energies and the non-detection of a SNR closeby, SNRs
are very unlikely to be a potential source for KM3-230213A.

\textbf{Stellar clusters:} Young stellar clusters with a large number of massive stars
can produce a collective wind termination shock where particles can be accelerated.
Inside the Mon R2 cloud there is ongoing star formation, but to a lesser extent than
in the Orion A and Orion B clouds. However, no Wolf-Rayet star was found in the region, and the most massive star is only \SI{\sim 10}{\solarmass} \citep{Rayner_2017_MonR2_star_formation}.
To search for potential young stellar clusters, we use the catalogue from \cite{Celli_2024_young_stellar_clusters}. The stellar clusters are displayed in \autoref{fig:sources} with magenta circles where the size of the circles corresponds to the cluster size. Two clusters are found within the KM3-230213A uncertainty region (FSR\_1117 and NGC\_2183), and three additional ones within \SI{5}{\degree} from the event (BDSB91, NGC\_2232, and vdBergh\_80). However, all of these clusters have collective wind powers $< \SI{e34}{\erg\per\second}$ and
therefore, there is not enough power within the collective stellar winds to allow
efficient CR acceleration to the required energies \citep{Morlino_2021_stellar_clusters, Mitchell_2024_stellar_clusters}. Given the low wind powers, also a scenario of a SNR exploding in a massive wind-blowing cluster as in \cite{Vieu_2023_SNRs_in_SFR} seems unlikely.

\textbf{X-ray binaries and microquasars:}
Microquasars are a relatively new source class in the very-high-energy $\gamma$-ray regime,
with SS433 being the first microquasar detected by HAWC and H.E.S.S.
\citep{HAWC_2018_SS433, HESS_2024_SS433}.
Recently, the LHAASO collaboration discovered $\gamma$-ray emission above
\SI{100}{\tera\electronvolt} from 5 sources \citep{LHAASO_2024_microquasars}.
However, microquasars might lack the potential to accelerate CRs to the energies needed.
In the model from \cite{Escobar_2022_microquasars_Emax}, for example, protons reach
maximum energies of only \SI{\sim e16}{\electronvolt}, which makes any
potential association with KM3-230213A unlikely.
X-ray binaries were searched in catalogues produced with eROSITA data
\citep{Neumann_2023_Erosita_X-ray-binaries, Avakyan_2023_Erosita_X-ray-binaries} and microquasars in the catalogue provided by \cite{Combi_2008_microquasar_candidates}. Only a subset of X-ray binaries
will also be a microquasar. As can be seen in \autoref{fig:sources},
no X-ray binary or microquasar is found close to KM3-230213A.

\textbf{Pulsars and pulsar wind nebulae:}
Pulsars and pulsar wind nebulae are efficient particle accelerators and prominent sources
in the $\gamma$-ray sky. A large number of LHAASO sources detected above
energies of \SI{100}{\tera\electronvolt} are potentially associated with high spin-down power
pulsars. From general arguments, the maximum achievable energies in units of
\si{\peta\electronvolt} can be written as
\citep{Wilhelmi_2022_pulsars_max_energy}
\begin{align}
    E_{\text{max}}(\si{\peta\electronvolt}) \approx 2 \eta_{\text e} \eta_{\text B}^{1/2} \dot E_{36}^{1/2},
    \label{eq:E_max_pulsar}
\end{align}
where $\eta_{\text e}$ is the ratio between the electric and magnetic field strength,
$\eta_{\text B}$ the fraction of the pulsar wind energy flux transferred to the magnetic
field, and $\dot E_{36}$ the spin-down power in units of \SI{e36}{\erg\per\second}.
In ideal conditions, $\eta_{\text e} \leq 1$, and $\eta_{\text B} \leq 1$ because of
energy conservation.

We searched for known pulsars in the ATNF pulsar catalogue \citep{Manchester_2005_ATNF_pulsar_catalog}. Only two pulsars are found within
a distance less than \SI{5}{\degree} from the neutrino event: B0559-05, with a distance
of \SI{4.4}{\degree}, and B0621-04 with a distance of \SI{3.8}{\degree}
(see \autoref{fig:sources}). They have spin-down
powers of \SI{8.3e32}{\erg\per\second} and \SI{2.9e31}{\erg\per\second}, respectively, which
is far too low to reach even \si{\peta\electronvolt} energies according to
\autoref{eq:E_max_pulsar}. While the situation can be different in other cases
such as in binary systems \citep{Bykov_2024_gamma-ray_binaries}, the energy loss rates
of the pulsars are orders of magnitude below what one might consider as a strong pulsar
and are not able to accelerate CRs to the required energies. We therefore conclude that no known pulsar can be a potential CR accelerator for
KM3-230213A.

If the cone of open field lines around a pulsar's magnetic pole does not point towards the Earth, no pulsed emission can be detected and a pulsar might not be recognised as such. However, a powerful pulsar is expected to produce a bright pulsar wind nebula, which could be visible in $\gamma$-rays. A large fraction of this electromagnetic radiation is believed to be of leptonic origin, but a hadronic component is not excluded. In the next section, a search for potential Galactic $\gamma$-ray sources is performed.

\section{Constraints from $\gamma$-ray observations}\label{sec:gamma-ray_neutrino_limits}
\subsection{Search for potential $\gamma$-ray counterparts}\label{sec:gamma-ray_counterparts}
    Neutrino production via hadronic collisions or photo-meson production
    inevitably also leads to the production of $\gamma$-rays.
    In the very-high-energy $\gamma$-ray region, data from
    HAWC, LHAASO, and Fermi-LAT are available. For a Galactic
    source accelerating particles up to more than \SI{100}{\tera\electronvolt},
    one would expect detections in both HAWC and LHAASO, unless the $\gamma$-rays are absorbed.

\begin{figure}
     \centering
     \includegraphics[width=\figurewidth]{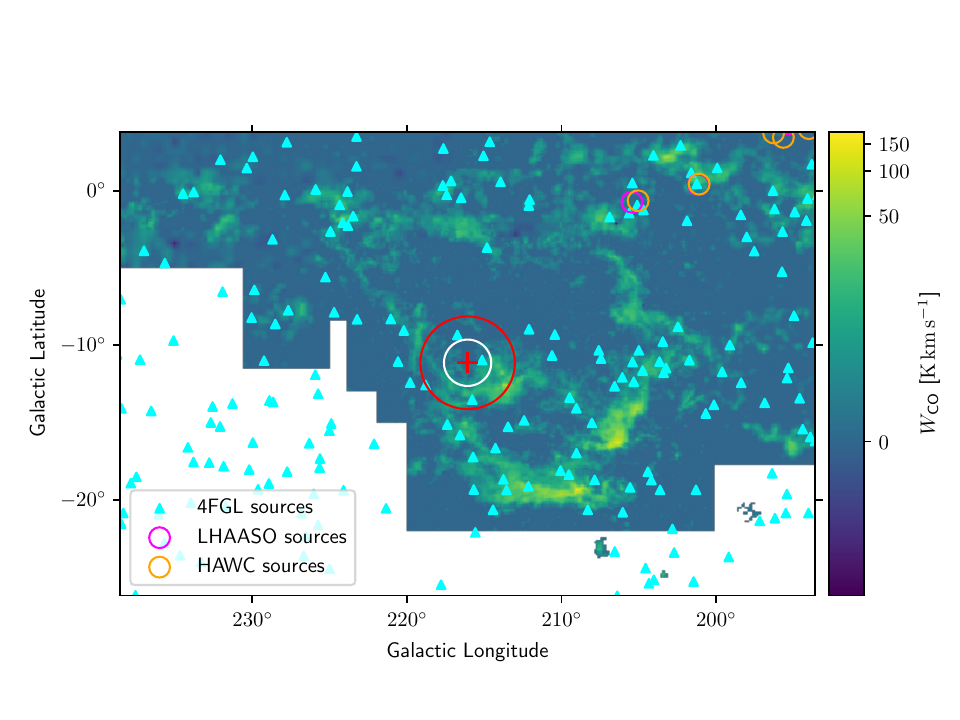}
     \caption{CO map from \cite{Dame_2001_CO_survey} of the velocity integrated brightness temperature tracing the molecular gas as a background, KM3-230213A position (red cross) and uncertainty (white and red circles),
     and sources from the 4FGL-DR4 catalogue \citep[cyan triangles,][]{Fermi_2023_4FGL-DR4}, the
     3HWC \citep[magenta circles,][]{HAWC_2020_3HWC_catalog} and the
     1LHAASO catalogue \citep[orange circles,][]{LHAASO_2024_first_catalogue}.}
     \label{fig:gamma-ray_sources}
    \end{figure}

    Fermi-LAT, HAWC and LHAASO are survey instruments that cover the
    region of interest. While Fermi-LAT observes $\gamma$-ray energies
    from tens of \si{\mega\electronvolt} up to several hundreds of
    \si{\giga\electronvolt}, the HAWC and LHAASO energy ranges extend up
    to hundreds of \si{\tera\electronvolt} and even \si{\peta\electronvolt}
    energies. Because of $\gamma$-ray absorption,
    at energies above hundreds of TeV, HAWC and LHAASO are effectively constrained
    to observations within the Milky Way, whereas at Fermi-LAT energies
    a large number of extragalactic objects are observed. We searched for
    potential counterparts
    in the 4FGL-DR4 Fermi-LAT catalogue \citep{Fermi_2023_4FGL-DR4}, the
    3HWC \citep{HAWC_2020_3HWC_catalog},
    and the first LHAASO catalogue \citep[1LHAASO,][]{LHAASO_2024_first_catalogue}
    of $\gamma$-ray sources. The
    4FGL-DR4, the 3HWC, and the 1LHAASO sources plotted over the CO map
    of \cite{Dame_2001_CO_survey} are shown in \autoref{fig:gamma-ray_sources}. The position of KM3-230213A and the \SI{68}{\percent} and \SI{99}{\percent} containment radii are also shown.

    No nearby sources from HAWC or LHAASO were found, imposing stringent
    constraints on any
    potential astrophysical sources in the region. While HAWC and LHAASO observations are much more relevant for the energy of the neutrino event, it is possible that the emission at HAWC and LHAASO energies is absorbed. This is different in the Fermi-LAT energy range, since even strong stellar ultraviolet radiation can not absorb \si{\giga\electronvolt} $\gamma$-rays and the electromagnetic cascades triggered by the absorption process can provide a relevant contribution to the emission at lower energies depending on the spectral shape. A source found by Fermi-LAT whose extension to higher energies is in tension with the non-detection by HAWC or LHAASO could be an absorbed Galactic source. Excluding extragalactic sources, there are two Fermi-LAT sources of unknown type within the
    \SI{99}{\percent} error region \citep{KM3NeT_2025_VHE_event, KM3NeT_2025_follow_up}:
    4FGL J0616.2-0653, and 4FGL J0624.8-0735.
    While 4FGL J0624.8-0735 has the potential radio source counterpart
    NVSS J062455-073536, of unknown type, and associated emission was detected in X-rays \citep{KM3NeT_2025_follow_up}, no potential counterpart at other wavelengths was found for 4FGL J0616.2-0653. As argued in \cite{KM3NeT_2025_follow_up}, the source might even be extended and its $\gamma$-ray emission could potentially be explained by mismodelled diffuse emission \citep{KM3NeT_2025_follow_up}.
    In addition to these catalogue sources, \cite{Marti_et_al_2013_MonR2_Fermi}
    performed an analysis of the source 2FGLJ0607.5-0618c, which is likely associated
    with the Mon R2 cloud. They reported no significant detection of $\gamma$-rays
    above energies of \SI{\sim 3}{\giga\electronvolt},
    while at lower energies the emission
    is consistent with a power-law $\text dN/\text dE \propto E^{-2.73}$.

    While no source in the vicinity of KM3-230213A was detected by HAWC or LHAASO,
    the HAWC collaboration published data accompanying the 3HWC catalogue
    \citep{HAWC_2020_3HWC_catalog}. These data include skymaps with the
    significances and flux normalisations maximising the likelihood
    function\footnote{The data are available under
    \url{https://data.hawc-observatory.org/datasets/3hwc-survey/fitsmaps.php}}.
    In this data, a point source and sources with extensions of
    \SI{0.5}{\degree}, \SI{1.0}{\degree}, and \SI{2.0}{\degree} are considered. The flux
    normalisations are given for an energy of \SI{7}{\tera\electronvolt}
    by assuming a power-law of $E^{-2.5}$.
    Limits on the $\gamma$-ray flux, computed using the reported flux and flux error, are used to derive a limit on the neutrino emission at the KM3-230213A location.
    
    Using the open-source
    \textit{gammapy} package \citep{gammapy:2023, gammapy:zenodo-1.2}, we retrieved the flux and
    uncertainty values, corresponding to the $\SI{2}{\sigma}$ Feldman-Cousins confidence intervals, from the skymaps.
    The values of the flux at
    \SI{7}{\tera\electronvolt}, the corresponding upper uncertainties
    and the derived upper limits (the sum of flux normalisation and uncertainty) for the point source and the extended
    source cases are shown in \autoref{tab:HAWC_limits}. As it can be seen, the uncertainties dominate over the flux normalisations since
    no HAWC source is located nearby. Using these upper limits on the flux
    at \SI{7}{\tera\electronvolt} and the power-law index of \num{-2.5},
    we computed the upper limits on the $\gamma$-ray
    flux for the whole HAWC energy range.
    Furthermore, the HAWC collaboration also has analysed the potential emission
    from nearby molecular clouds \citep{HAWC_2021_MonR2},
    amongst them Mon R2. While the cloud was not detected, upper limits
    were provided. These limits are also shown in the left panel of \autoref{fig:gamma-ray_and_neutrino_limits}, together with the prediction of the total emission from Mon R2 as a passive cloud from the same article.

    \begin{table*}
    \centering
    \begin{tabular}{|c|c|c|c|}
        \hline
        Source extension &
        Flux  at \SI{7}{\tera\electronvolt} [\si[]{\per\tera\electronvolt\per\square\centi\meter\per\second}] &
        Upper error [ \si[]{\per\tera\electronvolt\per\square\centi\meter\per\second}] &
        Upper limit [ \si[]{\per\tera\electronvolt\per\square\centi\meter\per\second}] \\
        \hline
        PS  & \num{9e-18} & \num{1.4e-16} & \num{1.5e-16} \\
        \hline
        \SI{0.5}{\degree} & \num{0} & \num{2.1e-16} & \num{2.1e-16} \\
        \hline
        \SI{1.0}{\degree} & \num{9e-17} & \num{4.2e-16} & \num{5.1e-16} \\
        \hline
        \SI{2.0}{\degree} & \num{1.6e-16} & \num{7.5e-16} & \num{9.1e-16} \\
        \hline
    \end{tabular}
    \caption{Flux normalisations from the public 3HWC data
    of HAWC \citep{HAWC_2020_3HWC_catalog} for KM3-230213A at
    \SI{7}{\tera\electronvolt} with the upper error (\SI{2}{\sigma} Feldman-Cousins confidence intervals) on the flux
    and the derived upper limits. The different rows show the
    point source (PS) case and the cases of extended sources with radii of
    \SI{0.5}{\degree}, \SI{1.0}{\degree}, and \SI{2.0}{\degree}.}
    \label{tab:HAWC_limits}
\end{table*}

    Since an inconsistency of the extension of the $\gamma$-ray spectra observed by Fermi-LAT with the upper limits obtained from HAWC data could be explained by an absorbed source potentially emitting neutrinos at higher energies, we compare in the left panel of \autoref{fig:gamma-ray_and_neutrino_limits} the Fermi-LAT spectra with the HAWC limits. An extension of the Fermi-LAT spectra towards higher energies does not necessarily represent the true spectral shape of the sources.
     As can be seen,
     the measurements of Mon R2 by Fermi-LAT are consistent with the upper
     limits of HAWC for Mon R2 at higher energies, assuming no spectral breaks or
     cutoffs. Similarly, the extrapolated fluxes of 4FGL J0616.2-0653 and
     4FGL J0624.8-0735 are well below the HAWC limits at the KM3-230213A
     position. However, the extension of the emission of Mon R2 from Fermi-LAT is below the predictions for Mon R2 at HAWC energies from \cite{HAWC_2021_MonR2}. A possible explanation for this discrepancy might be the use of the template for the Mon R2 cloud by HAWC, which was not done by \cite{Marti_et_al_2013_MonR2_Fermi}. Assuming a larger extension of the Mon R2 cloud will also yield larger $\gamma$-ray fluxes. Further uncertainties are the CR density, changes in the spectral shape and the particle density within Mon R2. Nevertheless, because the extensions of the Fermi-LAT spectra towards higher energies are consistent with the HAWC non-detection, the presence of a strong absorbed Galactic source is unlikely.

\subsection{Upper limits on neutrino fluxes from HAWC non-detections}
\label{sec:upper_limits_HAWC}

    With the upper limits from HAWC, one can derive limits on the maximum possible
    neutrino fluxes at the energy range of KM3-230213A.
    While HAWC can observe sources above
    energies of \SI{100}{\tera\electronvolt} \citep{HAWC_2020_UHE_sources}, it
    is not guaranteed that the assumption of a power-law index of \num{-2.5}
    is still valid at these energies and we therefore assumed that the
    limit only holds until \SI{100}{\tera\electronvolt}. This is the same
    maximum energy for which HAWC provided upper limits for Mon R2 in
    \cite{HAWC_2021_MonR2}.

    To obtain a limit on the neutrino flux at hundreds of \si{\peta\electronvolt},
    we made use of the publicly available GAMERA software package
    \citep{GAMERA_2015_ICRC, GAMERA_2022_ascl}. We assumed a population of
    power-law distributed CRs producing $\gamma$-rays
    and neutrinos via proton-proton collisions, fulfilling the HAWC limits.
    The most poorly constraining neutrino limits are obtained for hard CR power-law indices.
    Here, we choose a power-law index of the CRs of \num{-2}, which is the one expected
    in standard diffusive shock acceleration. Harder CR spectral indices are
    more difficult to produce, and we also note that none of the sources detected with
    the LHAASO Kilometer Squared Array (LHAASO KM2A,
    \cite{LHAASO_2024_first_catalogue}) has a power-law index harder
    than \num{-2}. Since for hadronic collisions the $\gamma$-ray and neutrino spectra
    show approximately the same power-law index as the parent CRs, the existence of
    Galactic sources with harder CR indices than \num{-2} above hundreds of
    \si{\tera\electronvolt} seems unlikely. For the fulfilment of the HAWC upper limits in $\gamma$-rays, this power-law index is assumed to be valid for energies above \SI{100}{\tera\electronvolt}.

    The $\gamma$-ray emission can be absorbed by
    the large-scale diffuse Galactic radiation fields or by radiation fields
    inside the source itself. To assess the impact of the large-scale Galactic
    radiation fields we made use of the radiation model developed by \cite{Popescu_2017}
    together with the cosmic microwave background. At \SI{100}{\tera\electronvolt},
    even if the neutrino would have been produced at the outer edge of the Milky
    Way at a Galactic radius of \SI{24}{\kilo\pc} (the maximum Galactic radius of
    the model from \cite{Popescu_2017}), only \SI{3.4}{\percent}
    of the $\gamma$-ray emission is lost. For the Mon R2 cloud, located at a distance of
    \SI{830}{\pc}, this value reduces to \SI{0.7}{\percent}. Therefore, one
    can safely neglect the absorption by the large-scale fields and consider the $\gamma$-ray upper limits as a proxy for the computation of the corresponding neutrino limits.

    \begin{figure}
        \centering
        \includegraphics[width=0.49\linewidth]{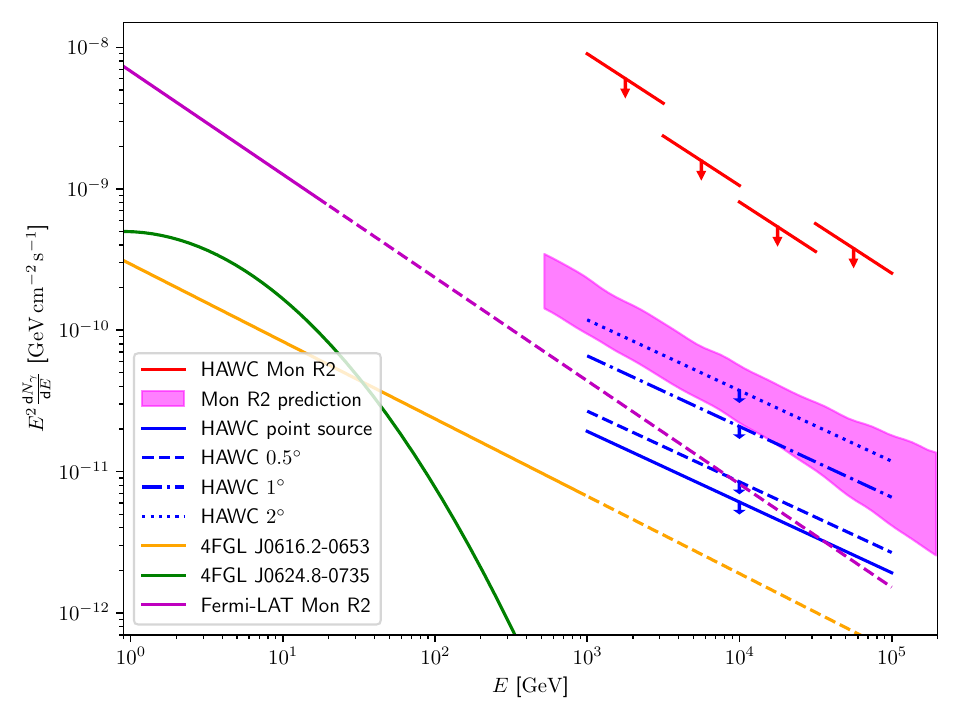}
        \includegraphics[width=0.49\linewidth]{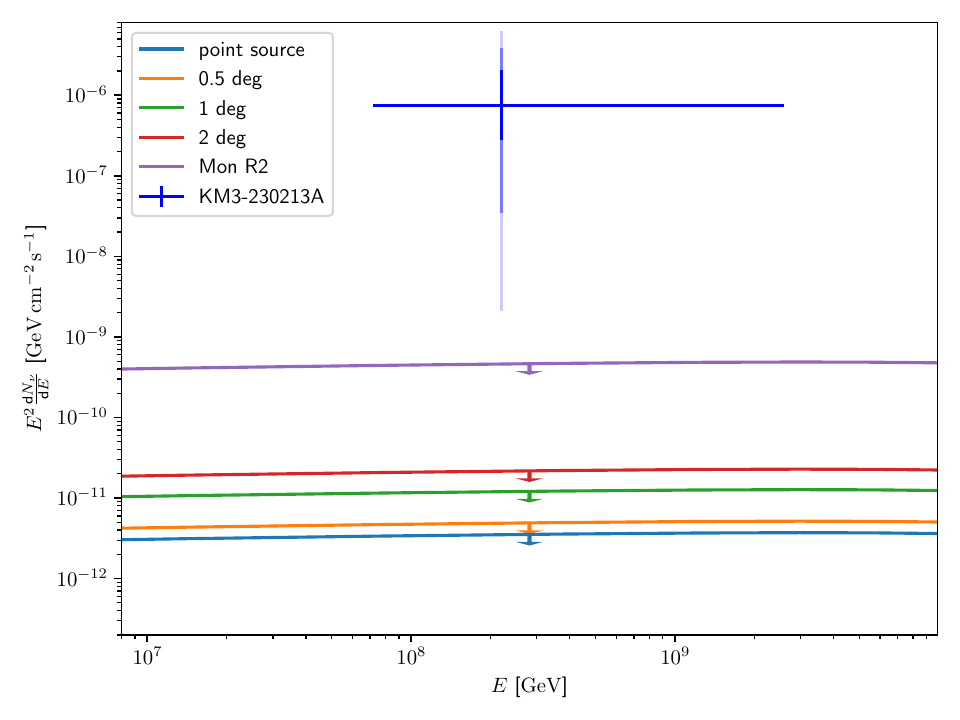}
        \caption{Left: Gamma-ray fluxes of the Fermi-LAT sources 4FGL J0616.2-0653,
     4FGL J0624.8-0735, and the Fermi-LAT spectrum of Mon R2 from
     \cite{Marti_et_al_2013_MonR2_Fermi} extended up to energies of
     \SI{100}{\tera\electronvolt} assuming the same spectral shape as at lower
     energies. The \SI{95}{\percent} credible interval
     upper limits on Mon R2 from HAWC are shown in red
     \citep{HAWC_2021_MonR2}, and the upper limits derived from
     public HAWC data for a point source and sources of \SI{0.5}{\degree},
     \SI{1}{\degree}, and \SI{2}{\degree} extension at the position of KM3-230213A are shown in blue. The prediction of the diffuse emission from Mon R2 as a passive cloud \citep{HAWC_2021_MonR2} is shown with the magenta band. Right: Neutrino limits derived from the HAWC $\gamma$-ray limits shown in the left panel
     for a point source, extended sources of \SI{0.5}{\degree}, \SI{1.0}{\degree},
     and \SI{2.0}{\degree}, and the Mon R2 cloud, assuming $E^{-2}$ distributed CRs. The point source
     flux derived from the detection of KM3-230213A is shown in blue, and the different
     shades of colour of the flux uncertainties represent the \SI{1}{\sigma}, \SI{2}{\sigma}, and
     \SI{3}{\sigma} Feldman-Cousins intervals.}
        \label{fig:gamma-ray_and_neutrino_limits}
    \end{figure}

    The derived neutrino limits for a point source,
    sources of \SI{0.5}{\degree}, \SI{1.0}{\degree},
    and \SI{2.0}{\degree} extension, and for the Mon R2 molecular cloud are shown in the right panel of \autoref{fig:gamma-ray_and_neutrino_limits}.
    Absorption was ignored, and it was assumed that the CRs follow an
    $E^{-2}$ power-law.
    The fluxes reported in \cite{KM3NeT_2025_VHE_event} were derived with the assumption of an all-sky diffuse isotropic flux. Such
    a flux can not be directly compared with the neutrino limits for individual sources
    with different extensions. We therefore computed a point-source flux for KM3-230213A for the current livetime of the detector from the effective area. For the calculation of the effective area, the same selection criteria as in \cite{KM3NeT_2025_VHE_event} were applied, and for the final result we also assumed a power-law spectral shape throughout the whole energy range
    from \SI{72}{\peta\electronvolt} to \SI{2.6}{\exa\electronvolt} with a spectral index
    of \num{-2} as in \cite{KM3NeT_2025_VHE_event}.
    This flux is also displayed in the right panel of \autoref{fig:gamma-ray_and_neutrino_limits}. All neutrino
    limits are more than three orders of magnitude below the expected flux, and the lower
    $\SI{3}{\sigma}$ uncertainty of the expected flux is still one order of magnitude above
    the limit of the Mon R2 cloud and two orders of magnitude or more for the
    other cases.
    Because the flux is derived from only one event, the flux derived from KM3-230213A can be considered as
    an upper limit to the neutrino flux. In this case, the limits derived from
    the HAWC data for a Galactic neutrino source are much more constraining.

    An underestimation of the neutrino limits can be caused by significant
    absorption within the source itself. The most relevant absorbing photon energy
    is $(m_{ \text e} c^2)^2/E_{\gamma}$, where $m_{\text e} c^2$ is the electron
    rest mass energy, and $E_{\gamma}$ the absorbed $\gamma$-ray energy
    \citep{Vernetto_Lipari_2016_absorption}.
    For $\gamma$-ray energies of \SI{100}{\tera\electronvolt}, the most relevant
    photons come from far-infrared radiation fields with energies around
    \SI{3}{\milli\electronvolt}. More accurate calculations with the open-source
    software package GAMERA yield temperature fields of \SI{60}{\kelvin}, corresponding to energies of \SI{5}{\milli\electronvolt}. Strong far-infrared radiation is produced in the Galaxy when surrounding clouds reprocess the light from massive stellar clusters. For non-ionizing ultraviolet radiation fields with energy densities of several hundreds of \si{\electronvolt\per\cubic\centi\meter}, one can produce far-infrared energy densities up to \SI{100}{\electronvolt\per\cubic\centi\meter} \citep{Dopita_et_al_2005, Groves_et_al_2008, Popescu_et_al_2011}. However, for a \SI{60}{\kelvin} radiation field with an energy density of \SI{100}{\electronvolt\per\cubic\centi\meter}, the attenuation length is \SI{\sim 400}{\pc}, which is much larger than the size of the whole Mon R2 cloud. It is therefore highly unlikely, that a strong hidden Galactic source producing large neutrino fluxes at the required energies exists.

\section{Conclusion}\label{sec:conclusion}
In this article, we explored the possibility that the highest energy candidate neutrino event ever detected
KM3-230213A could be produced within our own Galaxy. We searched for potential gas targets and found, that the Monoceros R2 cloud
is within the \SI{99}{\percent} containment radius and could serve as a
target for accelerated particles. However, the diffuse Galactic emission is insufficient to produce such an event flux. A search for potential $\gamma$-ray counterparts revealed two unidentified Fermi-LAT sources, but no detection by HAWC and LHAASO. Using publicly available HAWC data, we derived upper limits on the $\gamma$-ray and potential neutrino emission.
The limits derived from HAWC data on potential Galactic neutrino emission are orders of magnitude below the flux calculated from KM3-230213A when interpreted as a point-source flux.
The only hypothetical
exception could be an absorbed $\gamma$-ray source with a very hard spectrum. However,
nearly all conceivable Galactic CR accelerators struggle with producing the required
maximum energies and no counterpart can be found in the region.
We conclude that it is
very unlikely that the event originated from within the Milky Way.
Therefore, an extragalactic origin is much more likely.

\section*{Acknowledgements}
The authors acknowledge the financial support of:
KM3NeT-INFRADEV2 project, funded by the European Union Horizon Europe Research and Innovation Programme under grant agreement No 101079679;
Funds for Scientific Research (FRS-FNRS), Francqui foundation, BAEF foundation.
Czech Science Foundation (GAČR 24-12702S);
Agence Nationale de la Recherche (contract ANR-15-CE31-0020), Centre National de la Recherche Scientifique (CNRS), Commission Europ\'eenne (FEDER fund and Marie Curie Program), LabEx UnivEarthS (ANR-10-LABX-0023 and ANR-18-IDEX-0001), Paris \^Ile-de-France Region, Normandy Region (Alpha, Blue-waves and Neptune), France,
The Provence-Alpes-Côte d'Azur Delegation for Research and Innovation (DRARI), the Provence-Alpes-Côte d'Azur region, the Bouches-du-Rhône Departmental Council, the Metropolis of Aix-Marseille Provence and the City of Marseille through the CPER 2021-2027 NEUMED project,
The CNRS Institut National de Physique Nucléaire et de Physique des Particules (IN2P3);
Shota Rustaveli National Science Foundation of Georgia (SRNSFG, FR-22-13708), Georgia;
This work is part of the MuSES project which has received funding from the European Research Council (ERC) under the European Union’s Horizon 2020 Research and Innovation Programme (grant agreement No 101142396);
The General Secretariat of Research and Innovation (GSRI), Greece;
Istituto Nazionale di Fisica Nucleare (INFN) and Ministero dell’Universit{\`a} e della Ricerca (MUR), through PRIN 2022 program (Grant PANTHEON 2022E2J4RK, Next Generation EU) and PON R\&I program (Avviso n. 424 del 28 febbraio 2018, Progetto PACK-PIR01 00021), Italy; IDMAR project Po-Fesr Sicilian Region az. 1.5.1; A. De Benedittis, W. Idrissi Ibnsalih, M. Bendahman, A. Nayerhoda, G. Papalashvili, I. C. Rea, A. Simonelli have been supported by the Italian Ministero dell'Universit{\`a} e della Ricerca (MUR), Progetto CIR01 00021 (Avviso n. 2595 del 24 dicembre 2019); KM3NeT4RR MUR Project National Recovery and Resilience Plan (NRRP), Mission 4 Component 2 Investment 3.1, Funded by the European Union – NextGenerationEU,CUP I57G21000040001, Concession Decree MUR No. n. Prot. 123 del 21/06/2022;
Ministry of Higher Education, Scientific Research and Innovation, Morocco, and the Arab Fund for Economic and Social Development, Kuwait;
Nederlandse organisatie voor Wetenschappelijk Onderzoek (NWO), the Netherlands;
The grant “AstroCeNT: Particle Astrophysics Science and Technology Centre”, carried out within the International Research Agendas programme of the Foundation for Polish Science financed by the European Union under the European Regional Development Fund; The program: “Excellence initiative-research university” for the AGH University in Krakow; The ARTIQ project: UMO-2021/01/2/ST6/00004 and ARTIQ/0004/2021;
Ministry of Research, Innovation and Digitalisation, Romania;
Slovak Research and Development Agency under Contract No. APVV-22-0413; Ministry of Education, Research, Development and Youth of the Slovak Republic;
MCIN for PID2021-124591NB-C41, -C42, -C43 and PDC2023-145913-I00 funded by MCIN/AEI/10.13039/501100011033 and by “ERDF A way of making Europe”, for ASFAE/2022/014 and ASFAE/2022 /023 with funding from the EU NextGenerationEU (PRTR-C17.I01) and Generalitat Valenciana, for Grant AST22\_6.2 with funding from Consejer\'{\i}a de Universidad, Investigaci\'on e Innovaci\'on and Gobierno de Espa\~na and European Union - NextGenerationEU, for CSIC-INFRA23013 and for CNS2023-144099, Generalitat Valenciana for CIDEGENT/2018/034, /2019/043, /2020/049, /2021/23, for CIDEIG/2023/20, for CIPROM/2023/51 and for GRISOLIAP/2021/192 and EU for MSC/101025085, Spain;
Khalifa University internal grants (ESIG-2023-008, RIG-2023-070 and RIG-2024-047), United Arab Emirates;
The European Union's Horizon 2020 Research and Innovation Programme (ChETEC-INFRA - Project no. 101008324).

\bibliography{references}{}
\bibliographystyle{aasjournal}

\end{document}